\documentclass[aps,prd,twocolumn,superscriptaddress,floatfix,nofootinbib,showpacs,amsmath,amssymb,altaffilletter]{revtex4-1}
\pdfoutput=1

\UseRawInputEncoding

\usepackage{amsmath}
\usepackage{amsfonts}
\usepackage{amssymb}
\usepackage{graphicx}
\usepackage{bm}
\usepackage{subfigure}
\usepackage{url}
\usepackage[hyperindex]{hyperref}
\usepackage{color}
\usepackage[ddmmyy,24hr]{datetime}
\usepackage{bigdelim}
\usepackage{booktabs}
\usepackage{dcolumn}
\usepackage{multirow}
\usepackage{subfigure}
\usepackage{physics}
\usepackage{cancel}
\usepackage{stackrel}
\usepackage{paralist}
\usepackage{xspace}
\usepackage{slashed}
\usepackage{cancel}
\usepackage{todonotes}
\usepackage{enumerate}
\usepackage{float}
\usepackage{fullpage}
\usepackage{ulem}
\usepackage{wasysym}
\usepackage{comment}
\usepackage{bbold}
\usepackage{orcidlink}

\usepackage{feynmp}
\newcommand{\nua}[1]{\ensuremath{\rlap{\kern-2.5pt\ensuremath{\overset{\scriptscriptstyle(-)}{\phantom{\nu}}}}{\ensuremath{{\nu}_{#1}}}}}
\newcommand{\vet}[1]{\ensuremath{\hskip-1pt\vec{\hskip1pt#1}}}
\usepackage{enumitem}

\newcommand{\cenns}{CE$\nu$NS }

\begin{document}
\title{Reactor antineutrinos CE$\nu$NS on germanium: CONUS+ and TEXONO as a new gateway to SM and BSM physics}

\author{M. Atzori Corona \orcidlink{0000-0001-5092-3602}}
\email{mcorona@roma2.infn.it }
\affiliation{Istituto Nazionale di Fisica Nucleare (INFN), Sezione di Roma Tor Vergata, Via della Ricerca Scientifica, I-00133 Rome, Italy}

\author{M. Cadeddu \orcidlink{0000-0002-3974-1995}}
\email{matteo.cadeddu@ca.infn.it}
\affiliation{Istituto Nazionale di Fisica Nucleare (INFN), Sezione di Cagliari,
	Complesso Universitario di Monserrato - S.P. per Sestu Km 0.700,
	09042 Monserrato (Cagliari), Italy}

\author{N. Cargioli \orcidlink{0000-0002-6515-5850}}
\email{nicola.cargioli@ca.infn.it}
\affiliation{Istituto Nazionale di Fisica Nucleare (INFN), Sezione di Cagliari,
	Complesso Universitario di Monserrato - S.P. per Sestu Km 0.700,
	09042 Monserrato (Cagliari), Italy}

\author{F. Dordei \orcidlink{0000-0002-2571-5067}}
\email{francesca.dordei@cern.ch}
\affiliation{Istituto Nazionale di Fisica Nucleare (INFN), Sezione di Cagliari,
	Complesso Universitario di Monserrato - S.P. per Sestu Km 0.700,
	09042 Monserrato (Cagliari), Italy}

\author{C. Giunti \orcidlink{0000-0003-2281-4788}}
\email{carlo.giunti@to.infn.it}
\affiliation{Istituto Nazionale di Fisica Nucleare (INFN), Sezione di Torino, Via P. Giuria 1, I--10125 Torino, Italy}

\date{\dayofweekname{\day}{\month}{\year} \ddmmyydate\today, \currenttime}

\begin{abstract}
Coherent elastic neutrino-nucleus scattering (CE$\nu$NS) is a key process for probing Standard Model and beyond the Standard Model (BSM) properties. Following its first detection by the COHERENT Collaboration, recent reactor-based experiments provide a unique opportunity to refine our current understanding. In particular, the high-precision data from CONUS+, combined with the strong bounds from TEXONO, not only validate the CE$\nu$NS process at low energies but also provide improved constraints
on the weak mixing angle, neutrino electromagnetic properties (including the charge radius, millicharge, and magnetic moment), and nonstandard interactions and light mediators. We also examine the role of elastic neutrino-electron scattering, which gains significance in certain BSM scenarios and allows us to obtain the best limit for the millicharge of the electron neutrinos. By combining reactor and higher-energy spallation neutron source measurements, this work strengthens CE$\nu$NS as a precision tool for testing the Standard Model and beyond.
\end{abstract}

\maketitle  

\section{Introduction}

The coherent elastic neutrino-nucleus scattering (CE$\nu$NS) process, predicted by the Standard Model (SM) of particle physics~\cite{Freedman:1973yd}, represents a cornerstone in our understanding of neutrino interactions. First observed by the COHERENT Collaboration in 2017~\cite{COHERENT:2017ipa,COHERENT:2018imc} at the Spallation Neutron Source (SNS), \cenns occurs when a neutrino scatters off an entire nucleus coherently, leading to an enhancement in the interaction cross section that scales roughly quadratically with the number of neutrons~\cite{Cadeddu:2023tkp}. 
This process provides a unique tool for probing neutrino properties, nuclear structure, astrophysical and electroweak parameters and potential new physics beyond the Standard Model (BSM)~\cite{DeRomeri:2024hvc,Akimov:2024lnl,DeRomeri:2024iaw,Majumdar:2024dms,Pandey:2023arh,Coloma:2023ixt,AristizabalSierra:2024nwf, Cadeddu:2017etk, Cadeddu:2018dux, Cadeddu:2019eta, Cadeddu:2020lky, Cadeddu:2018izq, Cadeddu:2020nbr, Cadeddu:2021ijh, AtzoriCorona:2022moj,AtzoriCorona:2022qrf,AtzoriCorona:2023ktl,AtzoriCorona:2024rtv,Coloma:2017ncl,Liao:2017uzy,Lindner:2016wff,Giunti:2019xpr,Denton:2018xmq,AristizabalSierra:2018eqm,Miranda:2020tif,Banerjee:2021laz,Papoulias:2019lfi,Denton:2022nol,Papoulias:2017qdn,Dutta:2019nbn,Abdullah:2018ykz,Ge:2017mcq,Miranda:2021kre,Flores:2020lji, Farzan:2018gtr, Brdar:2018qqj}. 

Since then, significant progress has been made, and new observations have been achieved, making it possible for the investigation of \cenns signals to enter an exceptionally promising era.
Indeed, the initial observation was subsequently confirmed by an updated result with the same CsI detector~\cite{COHERENT:2021xmm} and other CE$\nu$NS detections, carried out by the COHERENT Collaboration in 2020 and 2024, using argon~\cite{COHERENT:2020iec} and germanium~\cite{COHERENT:2024axu} detectors, respectively. Moreover, the first indication of such a process using solar $^{8}$B neutrinos was also reported in 2024 by the dark matter XENONnT~\cite{XENON:2024ijk} and PandaX-4T Collaborations~\cite{PandaX:2024muv}.
In 2021, a \cenns observation from reactor antineutrinos was reported at the Dresden-II nuclear power plant using a germanium crystal detector~\cite{Colaresi:2021kus}. This detection relies on an increased value of the quenching factor~\cite{AtzoriCorona:2023ais}, necessary to translate the nuclear recoil energy into the measurable deposition in germanium, at low recoil energies. Namely, two quenching factor measurements were presented by the Dresden-II Collaboration which were in disagreement with the standard Lindhard theory prediction~\cite{Lindhard_theo} and other available measurements~\cite{SuperCDMS:2022nlc,CONUS:2020skt, Bonhomme:2022lcz, CONUS:2024wil, nGeN:2022uje}, which, however, were performed at higher recoil energies.
The situation has been clarified only recently.
First, TEXONO reported intriguing constraints on the \cenns cross section employing an electrocooled p-type point-contact germanium detector at the Kuo-Sheng Reactor Neutrino Laboratory~\cite{TEXONO:2024vfk} with a total exposure of 242~kg$\cdot$day, which appears to be consistent with the Lindhard model for the germanium quenching factor. Moreover, very recently, the CONUS Collaboration reported the first result obtained with their upgraded configuration, the CONUS+ experiment~\cite{Ackermann:2025obx}, which employs high-purity germanium crystal detectors with extremely low-energy thresholds of 160-180~eV installed at the nuclear power plant in Leibstadt, Switzerland. The full dataset, corresponding to a total exposure of 327~kg$\cdot$day, is consistent with the observation of a \cenns signal with a significance of 3.7$\sigma$ when using the Lindhard quenching model~\cite{Bonhomme:2022lcz}, in strong tension with the Dresden-II observation.\\

These results not only confirm the predicted quenching factor but also offer a new avenue for constraining SM and BSM phenomena. Indeed, the observation of \cenns with nuclear reactors offers unique advantages and complements studies at the SNS.
The lower energy of the antineutrinos from reactors (a few MeV) ensures that the nuclear form factor is effectively unity, making the results largely independent of nuclear physics uncertainties~\cite{Co:2020gwl}. This contrasts with SNS experiments, where higher-energy neutrinos introduce greater sensitivity to nuclear models. Reactor-based CE$\nu$NS experiments thus provide a clean, model-independent environment to study neutrino properties and nonstandard interactions. 
By combining the precise measurement from CONUS+ with the stringent limits from TEXONO, this paper aims to explore the implications of the recent \cenns searches at reactors. Whenever relevant, we also discuss the potential impact of including the COHERENT germanium data~\cite{COHERENT:2024axu} and the recent constraints from the $\nu$GEN Collaboration~\cite{nuGeN:2025mla},
which were released during the final stages of this work.
In particular, we provide updated measurements of the weak mixing angle and improved sensitivity to neutrino nonstandard properties~\cite{Giunti:2024gec}, like the charge radius, the electric millicharge, and magnetic moment, as well as to nonstandard interactions (NSIs) and the existence of new light mediators~\cite{AtzoriCorona:2022moj,Cadeddu:2020nbr}. We also discuss the contribution of elastic neutrino-electron scattering, which becomes notably significant when considering certain neutrino electromagnetic properties. 
Combined with the SNS, which probes higher-energy regimes, these studies create a comprehensive framework for exploring \cenns and testing the SM and BSM physics.

\section{Theoretical framework and data analysis}

In this section, the \cenns differential cross section in the SM is introduced.
Additionally, the phenomenology of elastic neutrino-electron scattering in the SM is briefly summarized. To conclude, we explain the data analysis techniques used in this paper.

\subsection{\cenns cross section}
The \cenns differential cross section as a function of nuclear recoil energy \(T_\mathrm{nr}\) for a neutrino \(\nu_\ell\) (\(\ell = e, \mu, \tau\)) scattering off a nucleus \(\mathcal{N}\) is given by
\begin{equation}
    \dfrac{d\sigma_{\nu_{\ell}\text{-}\mathcal{N}}}{d T_\mathrm{nr}} = 
    \dfrac{G_{\text{F}}^2 M}{\pi} 
    \left( 1 - \dfrac{M T_\mathrm{nr}}{2 E^2} \right)
    \left( Q^{V}_{\ell, \mathrm{SM}} \right)^2,
    \label{eq:cexsec}
\end{equation}
where \(G_{\text{F}}\) is the Fermi constant, \(E\) is the neutrino energy, \(M\) is the nuclear mass, and the weak nuclear charge is
\begin{equation}
    Q^{V}_{\ell, \mathrm{SM}} = \left[ g_{V}^{p}(\nu_\ell) Z F_Z(|\vec{q}|^2) + g_{V}^{n} N F_N(|\vec{q}|^2) \right].
    \label{eq:weakcharge}
\end{equation}
Here, \(Z\) and \(N\) are the numbers of protons and neutrons in the nucleus, respectively, and \(F_Z(|\vec{q}|^2)\), \(F_N(|\vec{q}|^2)\) are the nuclear form factors describing the loss of coherence at high momentum transfer \( |\vec{q}| \)~\cite{AtzoriCorona:2023ktl}. 

The coefficients $g_{V}^{n}$ and $g_{V}^{p}$ quantify the weak neutral-current interactions of neutrons and protons, respectively. In the SM, they correspond to 
\begin{align}
g_{V}^{p}(\nu_{e}) &= 0.0379,\, g_{V}^{p}(\nu_{\mu}) = 0.0297, \\
g_{V}^{p}(\nu_{\tau}) &= 0.0253,\, g_{V}^{n} = -0.5117, 
\end{align}
when taking into account radiative corrections in the $\overline{\mathrm{MS}}$ scheme~\cite{AtzoriCorona:2024rtv,AtzoriCorona:2023ktl,Erler:2013xha,PhysRevD.110.030001}. The small difference with respect to the coefficients found in Ref.~\cite{AtzoriCorona:2024rtv} is due to the new value of the weak mixing angle at low energies \(\sin^2 \theta_W = 0.23873\)~\cite{PhysRevD.110.030001}, which enters in the proton coefficients.   
 
For the isotopic composition of germanium, we use $(Z, N)_{\mathrm{{}^{70, 72, 73, 74, 76}Ge}} = (32, (38, 40, 41, 42, 44))$,
with the natural abundances $\textit{f}(^{70}\mathrm{Ge}) = 0.2057, \, \textit{f}(^{72}\mathrm{Ge}) = 0.2745,\, \textit{f}(^{73}\mathrm{Ge}) = 0.0775,\, \textit{f}(^{74}\mathrm{Ge}) = 0.3650,$ and $ \textit{f}(^{76}\mathrm{Ge}) = 0.0773$~\cite{BerglundWieser+2011+397+410}.

For COHERENT data, the form factors are essentially due to the higher energy of neutrinos, but in the energy range of the CONUS+ and TEXONO experiments, both proton and neutron form factors are practically unity. This simplifies the analysis, making the results independent of specific form factor parameterisations. We use the Helm parameterization for consistency, which is effectively equivalent to other common parameterisations~\cite{Helm:1956zz, Piekarewicz:2016vbn, Klein:1999qj}. The proton rms radii for Cs, I, Ar, and Ge nuclei are taken from spectroscopy and electron scattering data~\cite{Fricke:1995zz, Angeli:2013epw, Fricke2004}, while neutron radii are estimated from theoretical models~\cite{Hoferichter:2020osn, Cadeddu:2020lky,AtzoriCorona:2023ktl}. 

\subsection{Neutrino-electron elastic scattering}

Neutrino-electron elastic scattering ($\nu$ES) is a concurrent process to CE$\nu$NS. Within the SM, its contribution to the total event rate at low recoil energies is negligible and typically omitted in \cenns analyses. However, in BSM scenarios, the $\nu$ES contribution can increase significantly, making its inclusion important for obtaining stronger constraints~\cite{Coloma:2022avw}. Specifically, the effects of possible millicharges and magnetic moments are greatly amplified at low recoil energies.

For the COHERENT Ar dataset, the $f_{90}$ parameter~\cite{COHERENT:2020iec,COHERENT:2020ybo} enables efficient discrimination between nuclear recoils from \cenns and electron recoils from $\nu$ES, rendering the latter negligible. However, no such feature is available for the COHERENT CsI dataset or experiments like TEXONO and CONUS+, necessitating the inclusion of the $\nu$ES contribution in the analysis.

The SM cross section for neutrino-electron elastic scattering off an atom \(\mathcal{A}\) is
\begin{align} \nonumber
    \dfrac{d\sigma_{\nu_{\ell}-\mathcal{A}}}{d T_{\text{e}}}
    &=
    Z_{\text{eff}}^{\mathcal{A}}
    \,
    \dfrac{G_{\text{F}}^2 m_{e}}{2\pi}
    \left[ \vphantom{\left( 1 - \dfrac{T_{e}}{E} \right)^2}
    \left( g_{V}^{\nu_{\ell}} + g_{A}^{\nu_{\ell}} \right)^2
    +
    \left( g_{V}^{\nu_{\ell}} - g_{A}^{\nu_{\ell}} \right)^2
    \right.
    \\
    &
    \left.
    \left( 1 - \dfrac{T_{e}}{E} \right)^2
    -
    \left( (g_{V}^{\nu_{\ell}})^2 - (g_{A}^{\nu_{\ell}})^2 \right)
    \dfrac{m_{e} T_{e}}{E^2}
    \right],
    \label{eq:ES-cross-section}
\end{align}
where \(m_e\) is the electron mass, \(T_e\) is the electron recoil energy, and the flavor-dependent couplings are
\begin{align}
    g_{V}^{\nu_{e}} &= 0.9524, \quad g_{A}^{\nu_{e}} = 0.4938, \label{eq:gve} \\
    g_{V}^{\nu_{\mu}} &= -0.0394, \quad g_{A}^{\nu_{\mu,\tau}} = -0.5062, \label{eq:gvm}
\end{align}
when including also radiative corrections~\cite{AtzoriCorona:2022jeb,Erler:2013xha} and the latest weak mixing angle calculation~\cite{PhysRevD.110.030001}. For antineutrinos, \(g_{A} \to -g_{A}\).  

The factor $Z_{\text{eff}}^{\mathcal{A}}$  accounts for the number of electrons ionized at a given recoil energy \(T_e\), correcting the cross section derived under the Free Electron Approximation (FEA), which assumes free, stationary electrons~\cite{Kouzakov:2014lka}. Values of \(Z_{\text{eff}}^{\mathcal{A}}\) for Ge are provided in Table II of Ref.~\cite{AtzoriCorona:2022qrf}. 
In the sub-keV regime relevant for TEXONO and CONUS+, atomic effects become significant, requiring corrections to this approach. This can be achieved using \textit{ab initio} methods like the multiconfiguration relativistic random phase approximation (MCRRPA)~\cite{PhysRevA.25.634,PhysRevA.26.734,Chen:2013lba}, which better account for many-body dynamics. 
In the case of SM neutrino scattering or interactions with an additional neutrino magnetic moment (MM), the MCRRPA formalism results in a slight, nearly constant reduction of the expected $\nu$ES event rate as a function of the recoil energy.  

Conversely, when considering neutrino electric charges (ECs), the application of MCRRPA significantly enhances the low-energy electron-recoil spectrum in comparison to the FEA approach corrected for the stepping function. In this scenario, the equivalent photon approximation (EPA) can also be employed~\cite{Chen:2014ypv,Hsieh:2019hug}. This method links the ionization cross section to the photoabsorption one, effectively reproducing the MCRRPA prediction for a millicharged neutrino.
At higher energies, as in COHERENT CsI and Ar, the FEA corrected by \(Z_{\text{eff}}^{\mathcal{A}}\) provides a good approximation.

\subsection{Data analysis}

For the analysis of the COHERENT CsI and Ar data, we follow closely the strategy explained in detail in Refs.~\cite{AtzoriCorona:2023ktl,AtzoriCorona:2022qrf}. We obtained information on all the quantities used from Refs.~\cite{COHERENT:2020iec,COHERENT:2020ybo} for the Ar data and from Ref.~\cite{Akimov:2021dab} for the CsI data.

As a function of the electron-equivalent recoil energy, the theoretical \cenns  event rate $N^{\nu_{e}-\mathcal{N}}$ to be compared to the CONUS+~\cite{Ackermann:2025obx} and TEXONO~\cite{TEXONO:2024vfk} data is given by
\begin{align} \nonumber
    \dfrac{dN^{\nu_{e}-\mathcal{N}}}{dT_e} &=
N_T  \int_{T^{\prime\text{min}}_{\text{nr}}}^{T^{\prime\text{max}}_{\text{nr}}}  dT'_{\rm nr}\, \mathcal{R}(T_e,T'_e(T'_{\rm nr})) \\
&\int_{E_{\text{min}}(T'_{\text{nr}})}^{E_{\text{max}}} dE\, \dfrac{dN_\nu}{dE} \,
\dfrac{d\sigma_{\nu_{e}-\mathcal{N}}}{dT'_{\rm nr}},
\label{eq:rate}
\end{align}
where $N_T$ is the number of target germanium atoms per unit mass, $T'_{\mathrm{nr}(e)}$ is the true nuclear (electron) recoil kinetic energy, $T^{\prime\text{min}}_{\text{nr}} \simeq 2.96$~eV is the minimum average ionization energy in Ge, $T^{\prime\text{max}}_{\text{nr}} \simeq 2 E_{\text{max}}^2 / M$,
$E_{\text{max}} \sim 12 $~MeV, and
$E_{\text{min}}(T'_{\text{nr}}) \simeq \sqrt{MT'_\text{nr}/2}$.\footnote{This implies that the main contribution to \cenns for CONUS+ and TEXONO comes from antineutrinos with \mbox{$E\gtrsim 5-6\;\rm MeV$}, while for $\nu$ES the lower part of the energy spectrum is more relevant.} The energy resolution function $\mathcal{R}(T_e,T'_e)$ is defined by a Gaussian function
\begin{eqnarray}
    \mathcal{R}(T_e,T'_e)=\dfrac{1}{\sqrt{2\pi}\sigma_{\rm RES}}\,\mathrm{e}^{-\dfrac{(T_e-T'_e)^2}{2\sigma^2_{\rm RES}}},
\end{eqnarray}
where $T'_e=f_Q T'_{\rm nr}$, with $f_Q$ being the quenching factor. 
For CONUS+, $\sigma_{\rm RES}=\sqrt{\sigma_0^2+\mathcal{F_{\rm fano}}\epsilon_{\rm e-h} T_e'}$, where $\mathcal{F_{\rm fano}}=0.1096$ is the Fano factor for germanium, $\epsilon_{\rm e-h}=2.96\, \mathrm{eV}$ is the energy needed to create an energy-hole pair in germanium and $\sigma_0=20.38\, \mathrm{eV}$ comes from the measured $\mathrm{FWHM}=48\, \mathrm{eV}$~\cite{Ackermann:2025obx}, while for TEXONO the same applies but for $\sigma_0=29.8\, \mathrm{eV}$ corresponding to a $\mathrm{FWHM}=70.2\, \mathrm{eV}$~\cite{TEXONO:2024vfk}.
For both detectors, a standard Lindhard quenching factor is used with $k(\mathrm{Ge})=0.162\pm0.004$, as measured in Ref.~\cite{Bonhomme:2022lcz}. To derive the antineutrino flux, $dN_\nu/dE$, the neutrino spectra were constructed according to the prescription described in Ref.~\cite{Perisse:2023efm}, where a recent reevaluation of the so-called summation method for the prediction of the reactor antineutrino spectra is discussed. In this framework, the spectra are presented with a careful estimation of the uncertainty budget and align well with recent inverse beta decay measurements~\cite{An:2025ufv,STEREO:2022nzk}. The antineutrino emission is determined by the superposition of the energy spectra resulting from the fission of the four primary fission isotopes present in reactor fuel. These spectra are weighted according to the relative fission fractions characteristic of each nuclear power plant: 0.53, 0.08, 0.32, and 0.07 for $^{235}\rm U$, $^{238}\rm U$, $^{239}\rm Pu$, and $^{241}\rm Pu$ in the case of CONUS+~\cite{CONUS:2024lnu}, while for TEXONO the relative fission fractions are 0.55, 0.07, 0.32, and 0.06~\cite{TEXONO:2006xds}. On top of the neutrino flux produced by fission, an additional $\bar{\nu}_e$ contribution comes from the $\beta$ decay of activation products present in the fuel~\cite{Perisse:2023efm}. Then, for CONUS+ such spectra result in a flux $\Phi_{\rm{est}}=1.5\times 10^{13}\ \mathrm{cm^{-2} s^{-1}}$, considering a reactor power $P=3.6\ \mathrm{GW}_{\rm{th}}$ and a reactor-detector distance of $L=20.7\ \mathrm{m}$, which is in agreement with the antineutrino flux estimate reported in Refs.~\cite{Ackermann:2025obx,CONUS:2024lnu}. Similarly, for TEXONO we obtain $\Phi_{\rm{est}}=6.4\times 10^{12}\ \mathrm{cm^{-2} s^{-1}}$, that has been determined considering a reactor power $P=2.9\ \mathrm{GW}_{\rm{th}}$ and a reactor-detector distance of $L=28\ \mathrm{m}$~\cite{TEXONO:2006xds,TEXONO:2024vfk}.
Finally, in Eq.~(\ref{eq:rate}) the experimental acceptance does not appear since the data points provided are already corrected for it.\\

The theoretical \cenns event number $N^{\nu_{e}-\mathcal{N}}_{i}$ in each electron-recoil energy-bin $i$ is obtained by integrating the rate in Eq.~(\ref{eq:rate}) for a given electron-recoil energy range. 
We perform the analysis of the TEXONO Ge data using the least-squares function
\begin{eqnarray}\nonumber
    \chi^2_{\rm TEX}&=\sum_i \Big[\dfrac{N_i^{\rm exp}-\eta N_i^{\rm th}-\beta}{\sigma_i}\Big]^2+\\
    &+\Big(\dfrac{\beta-\beta_{\rm ^{135}Xe}}{\sigma_\beta}\Big)^2+\Big(\dfrac{\eta-1}{\sigma_\eta}\Big)^2,\label{eq:chi-texono}
\end{eqnarray}
where $N_i^{\rm exp}$ is the number of events observed experimentally in the $i$th bin, $N_i^{\rm th}$ is the predicted number of events evaluated in the physic scenario under consideration by modifying accordingly the \cenns cross section, and 
$\sigma_i$ is the statistical uncertainty on the $i$th number of observed events. The nuisance parameter $\eta$ takes into account the systematic uncertainty due to the neutrino flux and quenching factor, $\sigma_\eta=0.05$, and $\beta_{\rm ^{135}Xe}=1.71\, \mathrm{keV^{-1} kg^{-1} day^{-1}}$ is the number of events due to the Compton $^{135}$Xe background, with uncertainty $\sigma_\beta=0.35\, \mathrm{keV^{-1} kg^{-1}day^{-1}}$~\cite{TEXONO:2024vfk}.

We perform the analysis of the CONUS+ Ge data using the least-squares function
\begin{eqnarray}
    \chi^2_{\rm CON+}&=\sum_i \Big[\dfrac{N_i^{\rm exp}-\eta N_i^{\rm th}}{\sigma_i}\Big]^2+
    \Big(\dfrac{\eta-1}{\sigma_\eta}\Big)^2,\label{eq:chi-c+}
\end{eqnarray}
where $\sigma_\eta=0.17$ is the systematic uncertainty due to the neutrino flux, threshold, and quenching factor~\cite{CONUS:2024lnu, private}. No background contribution is present, as for CONUS+ we consider the residual signal distribution after the background subtraction.\\

\begin{figure}[t]
    \centering
        \subfigure[]{\label{fig:rate_texono}
    	\includegraphics[width=0.95\linewidth]{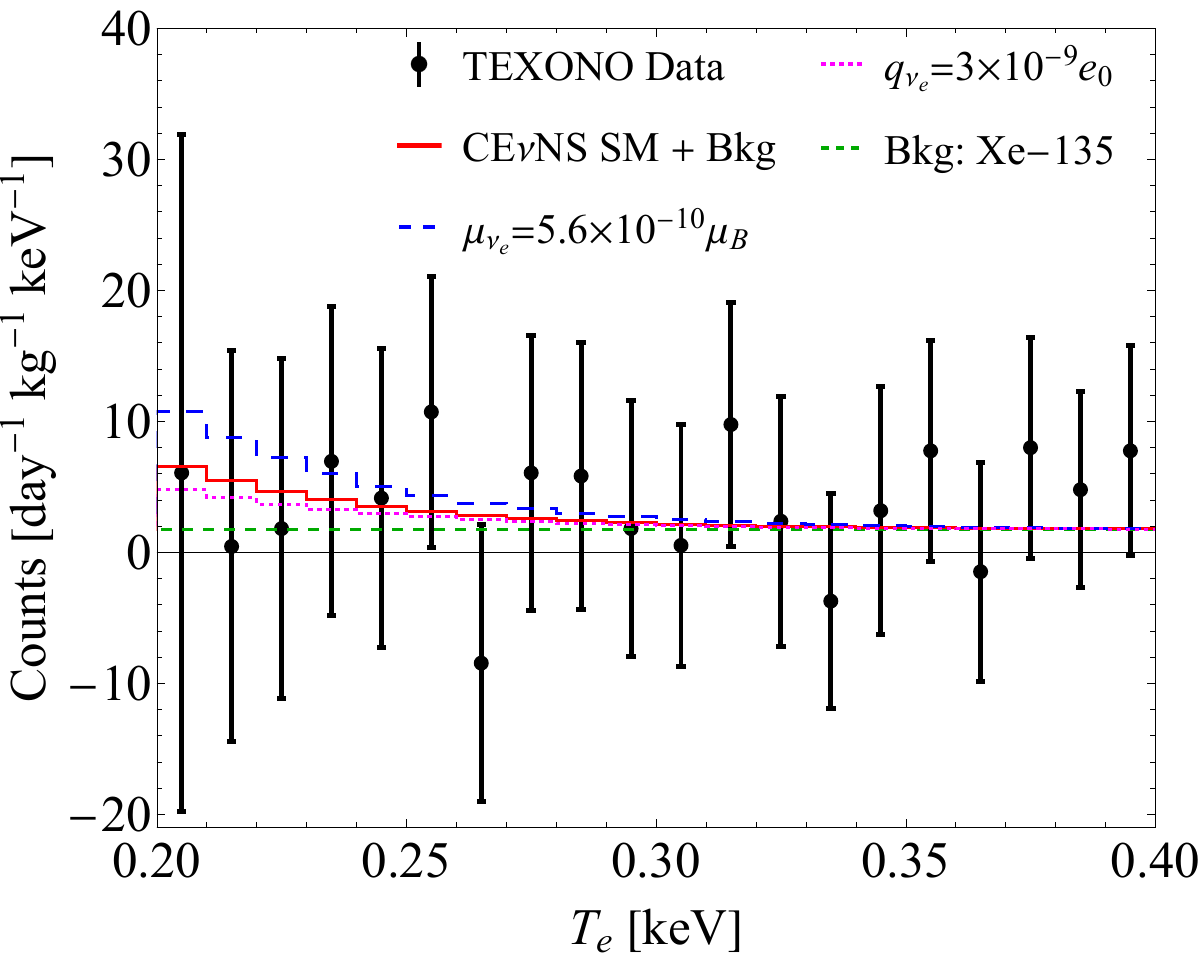}}
    	\subfigure[]{\label{fig:rate_conus}
    	\includegraphics[width=0.95\linewidth]{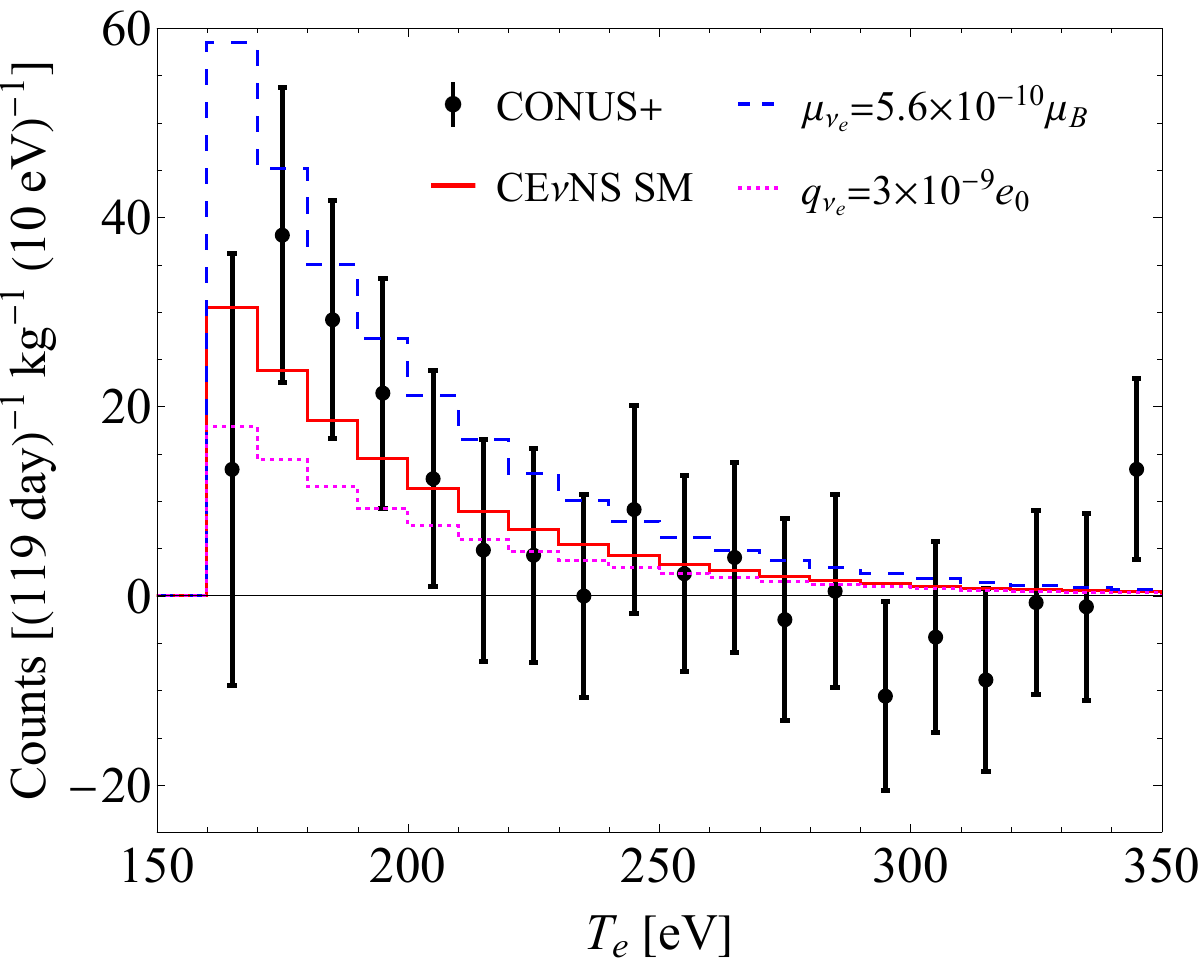}}
    \caption{(a) TEXONO and (b) CONUS+ data along with the SM \cenns theoretical prediction (solid red line) and those in the presence of a possible neutrino magnetic moment $\mu_{\nu_{e}}=5.6\times10^{-10}~\mu_{\text{B}}$ (dashed blue line), and a neutrino millicharge $q_{\nu_{e}}=3\times10^{-9}~e_0$ (dotted pink line). TEXONO data and predictions include also the background contribution due to Compton $^{135}$Xe (dashed green).}  
    \label{fig:rate_texono_conus}
\end{figure}

In Figs.~\ref{fig:rate_texono} and~\ref{fig:rate_conus} we show the TEXONO~\cite{TEXONO:2024vfk} and CONUS+~\cite{CONUS:2024lnu} (reported by the Collaboration in the form an effective single detector of 1~kg with a threshold of 160~eV) data along with the \cenns predictions, respectively, under different hypotheses.
In this way, one can compare the SM \cenns prediction, with those obtained in the presence of a possible neutrino magnetic moment, considering, e.g., $\mu_{\nu_{e}}=5.6\times10^{-10}~\mu_{\text{B}}$, and a possible neutrino millicharge of $q_{\nu_{e}}=3\times10^{-9}~e_0$. In these BSM scenarios, it is convenient to consider also the contribution due to the neutrino-electron elastic scattering, given that the $\nu$ES process is very sensitive to these quantities. Therefore, in such scenarios, the number of predicted events, $N_i^{\rm th}$, inside Eqs.~(\ref{eq:chi-texono}) and~(\ref{eq:chi-c+}) includes also the events due to the $\nu$ES process, obtained by properly modifying the rate definition in Eq.~(\ref{eq:rate}). \\

\begin{figure}[t]
    \centering
    \includegraphics[width=0.95\linewidth]{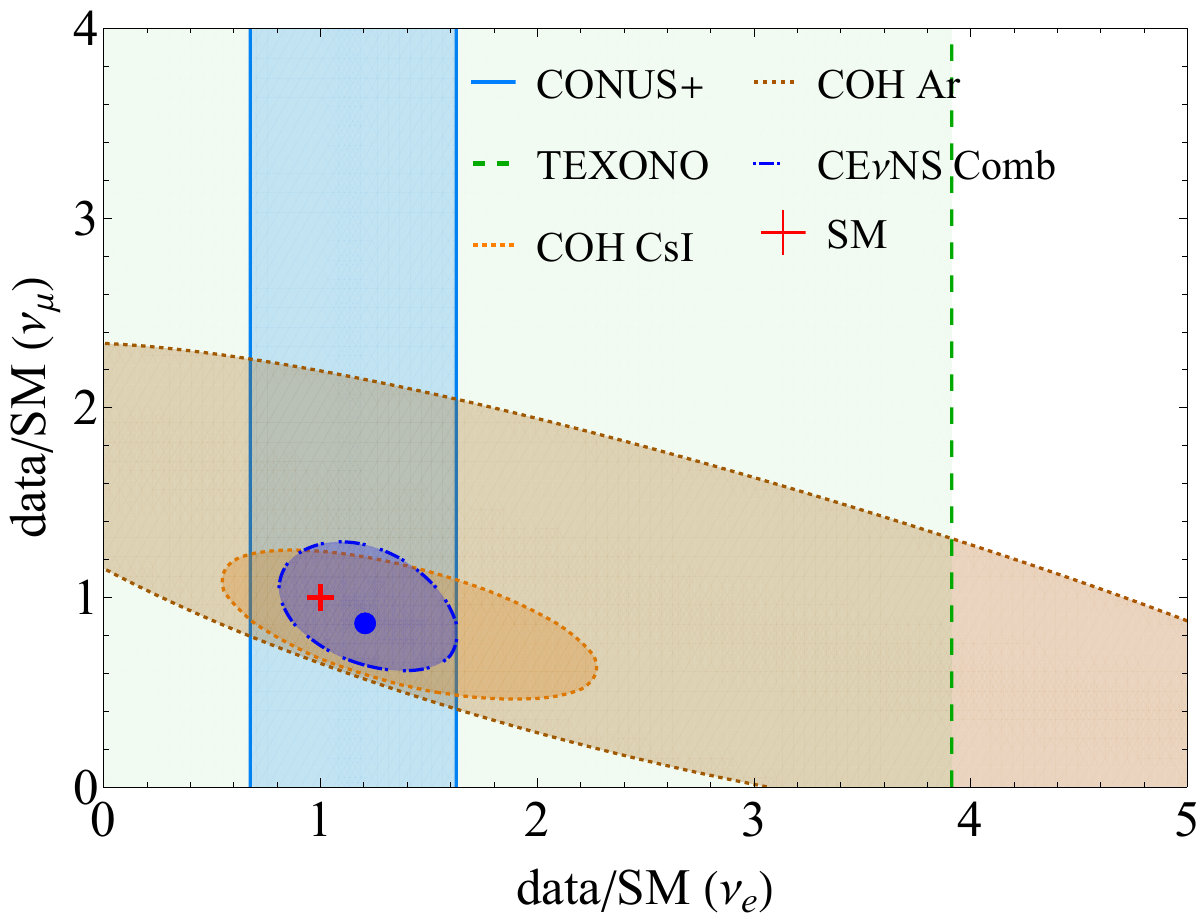}
    \caption{Level of agreement between \cenns data and SM predictions for TEXONO (green area), CONUS+ (light blue area) and COHERENT CsI (orange area) and Ar (brown area) together with their combination (blue area) and best fit (blue point) divided for electron and muon neutrinos. Contours are shown at 1$\sigma$ confidence level and the red cross indicates the perfect agreement with the SM.}
    \label{fig:eta}
\end{figure}

As visible in Figs.~\ref{fig:rate_texono} and~\ref{fig:rate_conus}, data and predictions show very good agreement, which can be quantified by looking at the $\eta$=data/SM nuisance parameter of the corresponding least-square functions in the SM case for TEXONO, CONUS+ and COHERENT CsI and Ar. 
This analysis results in $\eta=1.15\pm0.32$ and $\eta<4.2~(90\%\, \mathrm{C.L.})$ for CONUS+ and TEXONO, respectively, which well agrees with the values reported in Refs.~\cite{Ackermann:2025obx,TEXONO:2024vfk}.
The latter can be also split according to the neutrino flavor. The results are shown in Fig.~\ref{fig:eta} along with their combination, displaying a very good agreement with the SM. Nevertheless, the current precision allows us to put meaningful constraints on different SM and BSM scenarios, which will be analyzed in the next section.

\section{Results}

In this section, we present updated measurements and constraints obtained with the CONUS+ and TEXONO data and we compare and, whenever meaningful, combine them with SNS results. Recently, other manuscripts~\cite{Alpizar-Venegas:2025wor,Chattaraj:2025fvx,DeRomeri:2025csu} appeared which also analyzed the CONUS+ data. The results presented here, when a comparison is possible, show good agreement with them, also due to the relatively large statistical uncertainties, particularly in some parameters. In general, we note that we use a different treatment of the CONUS+ data as well as of the signal prediction with respect to Refs.~\cite{Alpizar-Venegas:2025wor,Chattaraj:2025fvx} considering an effective detector of 1~kg with a threshold of 160~eV, and we use a smaller systematic uncertainty following the prescriptions of the CONUS+ Collaboration~\cite{CONUS:2024lnu, private}.\\
We finally checked that the impact of the usage of different antineutrino flux models~\cite{Mueller:2011nm, Estienne:2019ujo, Vogel:1989iv, Kopeikin:1999tc,Kopeikin:2012zz} is minimal and does not change the results obtained.

\subsection{Weak mixing angle}

The weak mixing angle, $\theta_{\text{W}}$, is a key parameter in the electroweak interaction theory. It has been measured across various energy scales~\cite{PhysRevD.110.030001}, as its value can be significantly altered in certain BSM scenarios~\cite{Cadeddu:2021dqx}. Notably, low-energy determinations of $\theta_{\text{W}}$ play a crucial and complementary role to high-energy measurements, offering high sensitivity to extra $Z$ ($Z'$) bosons predicted by grand unified theories, technicolor models, supersymmetry, and string theories~\cite{Safronova_2018,Cadeddu:2021dqx,Corona:2021yfd,Cadeddu:2024baq}. This highlights the importance of improving experimental precision in the low-energy regime, where current measurements still exhibit substantial uncertainties.

\begin{figure}[t]
    \centering
    \includegraphics[width=0.95\linewidth]{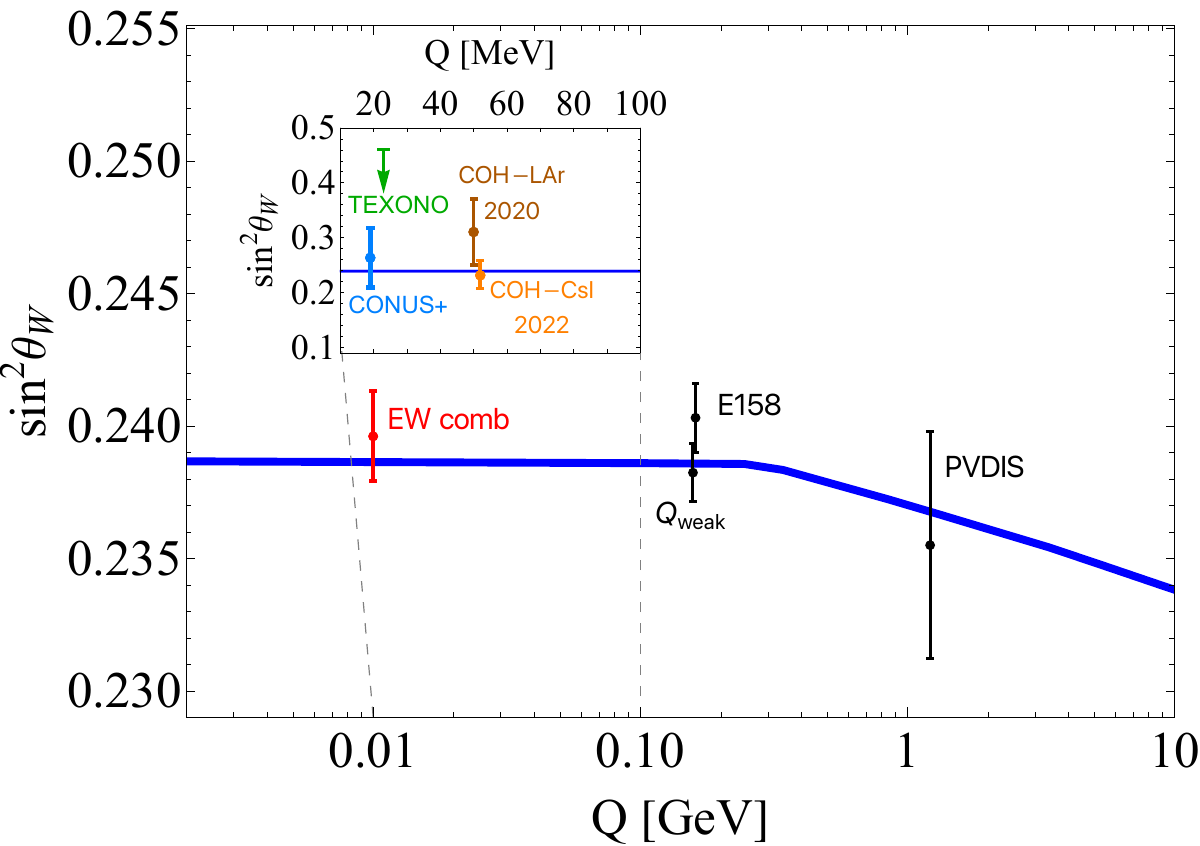}
    \caption{Variation of $\sin^2 \theta_{\text{W}}$ with the energy scale $Q$ in the low-energy range. The SM prediction is shown as the blue solid curve, together with experimental determinations in black from M{\o}ller scattering (E158)~\cite{Anthony:2005pm}, deep inelastic scattering of polarized electrons on deuterons (PVDIS)~\cite{Wang:2014bba}, and the result from the proton's weak charge ($ \text{Q}_{\text{weak}} $)~\cite{Androic:2018kni} and in red from an electroweak combined fit~\cite{AtzoriCorona:2024vhj}. The result derived in this paper for CONUS+ data is shown in blue in the inset, together with the CE$\nu$NS-only CsI~\cite{AtzoriCorona:2023ktl} and Ar~\cite{Cadeddu:2020lky} COHERENT determinations. The green arrow indicates the $1\sigma$ C.L. upper bound also obtained in this paper for the TEXONO data.}
    \label{fig:sin2}
\end{figure}

As shown in Ref.~\cite{AtzoriCorona:2023ktl}, the uncertainty obtained for the weak mixing angle from the CsI COHERENT dataset as well as that obtained with Ar~\cite{Cadeddu:2020lky} is still rather large when compared to the other determinations at low-momentum transfer, as visible also in the inset of Fig.~\ref{fig:sin2}. Furthermore, as demonstrated in Refs.~\cite{Cadeddu:2021ijh,AtzoriCorona:2024vhj,Cadeddu:2018izq}, the determination of the COHERENT weak mixing angle is highly sensitive to the choice of the poorly known neutron distribution radii. This strong correlation necessitates a simultaneous fit of these parameters to achieve a model-independent measurement of $\sin^2{\theta_{\text{W}}}$. By performing a combined global analysis of all available electroweak probes, it has been possible to place stringent constraints on the weak mixing angle while properly accounting for its correlation with the nuclear sector~\cite{AtzoriCorona:2024vhj}, as shown by the red point in Fig.~\ref{fig:sin2}.

On the contrary, in the analysis of the CONUS+ and TEXONO data the form factor of both protons and neutrons is practically equal to unity, making the particular choice of the value of the Ge neutron radius completely irrelevant. 
Here, we show the result of the fit of the CONUS+ and TEXONO \cenns data
\begin{align}
\sin^2\theta_W(\mathrm{CONUS+})&=0.26\pm0.05\,(1\sigma), \\ 
\sin^2\theta_W(\mathrm{TEXONO})&<0.46\, (1\sigma),
\end{align}
which are also shown in Fig.~\ref{fig:sin2}, compared to the other available measurements and the predicted variation of $\sin^2 \theta_{\text{W}}$ with the energy scale $Q$ in the low-energy range~\cite{PhysRevD.110.030001}. Good agreement is present, although the still large uncertainties must be taken into account. 
Here, we note that the precision of the weak mixing angle measurement is strongly affected by systematic uncertainties on the CE$\nu$NS signal. Therefore, any improvement in the characterization of the flux and detector response would clearly enhance the overall precision of the measurement. Moreover, we expect that the COHERENT Ge dataset will provide a precision slightly better than that obtained with COHERENT Ar, whereas $\nu$GEN data are expected to reach a level of precision comparable to TEXONO.

\subsection{Neutrino charge radii}

In the SM, neutrino charge radii (CR) are the only nonzero electromagnetic properties of neutrinos, appearing as radiative corrections to $g_{V}^{p}(\nu_{\ell})$~\cite{Giunti:2024gec}. The SM neutrino CR are given by~\cite{Bernabeu:2000hf,Bernabeu:2002nw}
\begin{equation}
\langle{r}_{\nu_{\ell}}^2\rangle_{\text{SM}}
=
-
\frac{G_{\text{F}}}{2\sqrt{2}\pi^2}
\left[
3 - 2 \ln\left(\frac{m_{\ell}^2}{m_W^2}\right)
\right],
\label{eq:cr-sm}
\end{equation}
where $m_{W}$ is the $W$ boson mass and $m_{\ell}$ is the mass of the charged lepton $\ell = e, \mu, \tau$. 
Numerically, the SM predictions are
\begin{align}
\langle{r}_{\nu_{e}}^2\rangle_{\text{SM}} &= -0.83 \times 10^{-32} \, \text{cm}^2, \label{eq:cr-e} \\
\langle{r}_{\nu_{\mu}}^2\rangle_{\text{SM}} &= -0.48 \times 10^{-32} \, \text{cm}^2. \label{eq:cr-mu}
\end{align}

Beyond the SM, neutrino CR may include off-diagonal terms (transition CR), $\langle r_{\nu_{\ell\ell'}}^2 \rangle$, in the flavor basis, generated by BSM effects~\cite{Kouzakov:2017hbc,AtzoriCorona:2022qrf,Cadeddu:2018dux}. However, here we will only consider the scenario including diagonal terms. To make evident the dependence of the \cenns cross section on the neutrino charge radii, we can substitute the neutrino-proton coupling in Eq.~(\ref{eq:weakcharge}) with $g_V^p= \tilde{g}_{V}^{p} - \tilde{Q}_{\ell}$,
where $\tilde{g}_{V}^{p}=0.0182$ is the neutrino-proton coupling without CR contributions and $\tilde{Q}_{\ell}$ represents the CR contribution
\begin{equation}
\tilde{Q}_{\ell} = \dfrac{\sqrt{2} \pi \alpha}{3 G_{\text{F}}} \langle r_{\nu_{\ell}}^2 \rangle,
\label{eq:cr-term}
\end{equation}
with $\alpha$ the fine-structure constant.
\begin{figure}[t]
    \centering
    \includegraphics[width=0.95\linewidth]{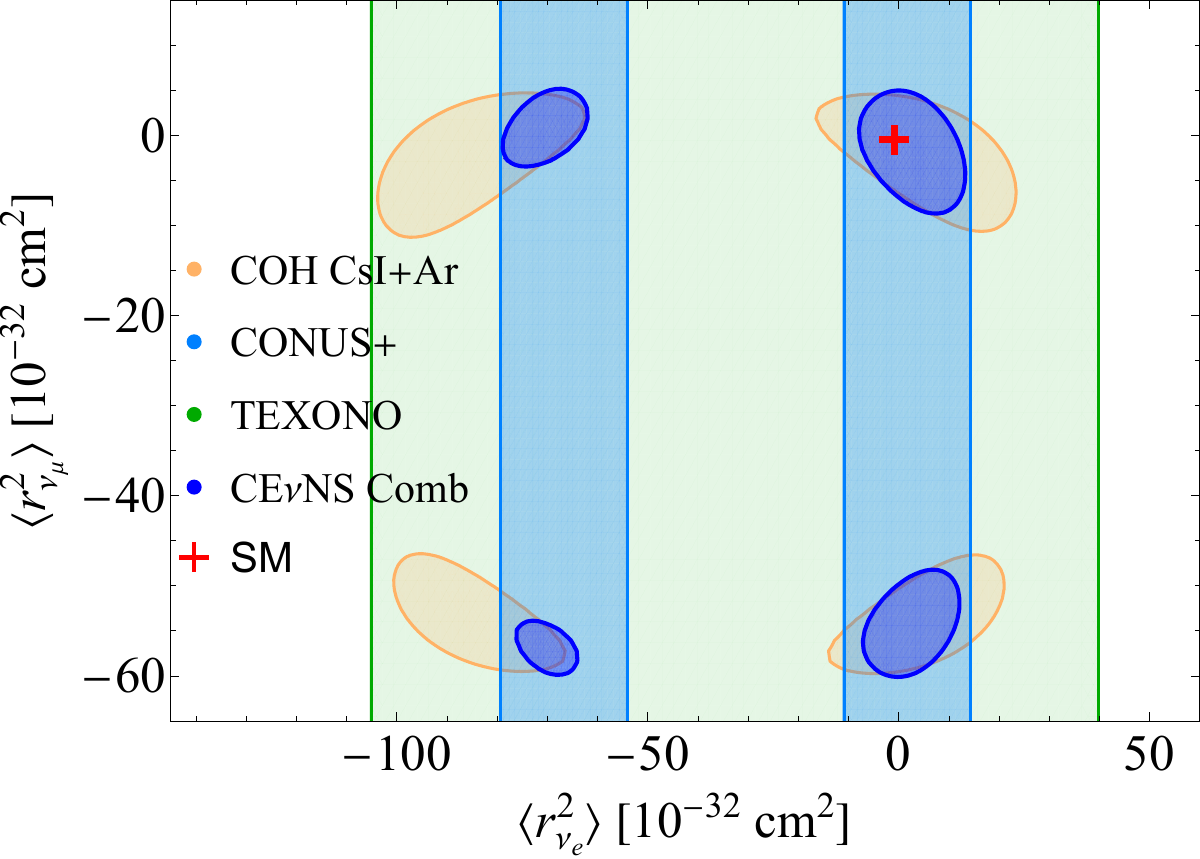}    
    \caption{Allowed regions at 90\% C.L. from the analysis of the latest COHERENT CsI and Ar data (COH), CONUS+, TEXONO and their combination (\cenns Comb) in the $\langle r^2_{\nu_{e}}\rangle$ vs $\langle r^2_{\nu_{\mu}}\rangle$ plane in case of a momentum-dependent neutrino charge radii correction. The red cross indicates the SM values in Eqs.~(\ref{eq:cr-e}) and~(\ref{eq:cr-mu}).
    }
    \label{fig:neutrinoCR}
\end{figure}

Reactor experiments like TEXONO and CONUS+, which only depend on the electron neutrino flavor, provide complementary information to higher-energy setups like COHERENT, where nuclear effects become more significant.
Moreover, as discussed in Ref.~\cite{AtzoriCorona:2024rtv}, the CR radiative correction shows a mild dependence on the momentum transfer. Given that reactor \cenns experiments are performed at lower energies with respect to COHERENT, the impact of such a momentum dependence is significantly less relevant and a combination of all available datasets allows to significantly reduce the available phase space. In particular, using the \cenns channel only we derive the results reported in Fig.~\ref{fig:neutrinoCR}. 

The numerical results obtained from our analysis on $\langle r^2_{\nu_{e}}\rangle$ at 90\% C.L. for CONUS+ and TEXONO are 
\begin{align}
    \langle{r_{\nu_{e}}^{2}}\rangle\,[10^{-32}\, \mathrm{cm}^2] &\in [-76,-57] \wedge [-8,11],\\
    \langle{r_{\nu_{e}}^{2}}\rangle\,[10^{-32}\, \mathrm{cm}^2] &\in [-98,32],
    \label{eq:CR_limit}
\end{align}
respectively.
The combined result for $\langle r^2_{\nu_{e}}\rangle$ and $\langle r^2_{\nu_{\mu}}\rangle$, when including also COHERENT CsI and Ar measurements~\cite{AtzoriCorona:2024rtv} are 
\begin{align}
    \langle{r_{\nu_{e}}^{2}}\rangle\,[10^{-32}\, \mathrm{cm}^2] &\in [-73,-67] \wedge [-5,11],\\
    \langle{r_{\nu_{\mu}}^{2}}\rangle\,[10^{-32}\, \mathrm{cm}^2] &\in [-58,-50] \wedge [-7,3].
    \label{eq:CR_limit_comb}
\end{align}
For a comparison with other measurements, see Ref.~\cite{Giunti:2024gec}. We expect that the $\nu$GEN constraint is similar to that achieved with TEXONO data. In contrast, the COHERENT germanium data could play a more relevant role due to their sensitivity to both muon and electron neutrino flavors. However, given the limited statistics collected so far, the dominant contribution is still expected to come from the COHERENT CsI dataset.

\subsection{Neutrino magnetic moment}

The neutrino magnetic moment is the most investigated neutrino electromagnetic
property, both theoretically and experimentally.
Indeed, its existence is predicted by many BSM theories,
especially those that include right-handed neutrinos, see the reviews in Refs.~\cite{Giunti:2014ixa,Giunti:2015gga,Giunti:2024gec}.
The MM contribution does not interfere with the SM one, and thus it is accounted for by adding to the SM cross section in Eq.~(\ref{eq:cexsec}) the MM contribution, namely
\begin{equation}
\dfrac{d\sigma_{\nu_{\ell}\text{-}\mathcal{N}}^{\text{MM}}}{d T_\mathrm{nr}}
=
\dfrac{ \pi \alpha^2 }{ m_{e}^2 }
\left( \dfrac{1}{T_\mathrm{nr}} - \dfrac{1}{E} \right)
Z^2 F_{Z}^2(|\vec{q}|^2)
\left| \dfrac{\mu_{\nu_{\ell}}}{\mu_{\text{B}}} \right|^2
,
\label{cs-mag}
\end{equation}
where $\mu_{\nu_{\ell}}$ is the effective MM of the flavor neutrino $\nu_{\ell}$~\cite{Giunti:2014ixa}
and $\mu_{\text{B}}$ is the Bohr magneton.

\noindent In the case of neutrino-electron scattering, the cross section in the presence of neutrino magnetic moments receives as well an additional contribution equal to
\begin{equation}
\dfrac{d\sigma_{\nu_{\ell}\text{-}\mathcal{A}}^{\text{MM}}}{d T_\mathrm{e}}
=
Z_{\text{eff}}^{\mathcal{A}}(T_{\text{e}}) \dfrac{ \pi \alpha^2 }{ m_{e}^2 }
\left( \dfrac{1}{T_\mathrm{e}} - \dfrac{1}{E} \right)
\left| \dfrac{\mu_{\nu_{\ell}}}{\mu_{\text{B}}} \right|^2.
\label{es-mag}
\end{equation}

\begin{figure}[t]
    \centering
    \includegraphics[width=0.95\linewidth]{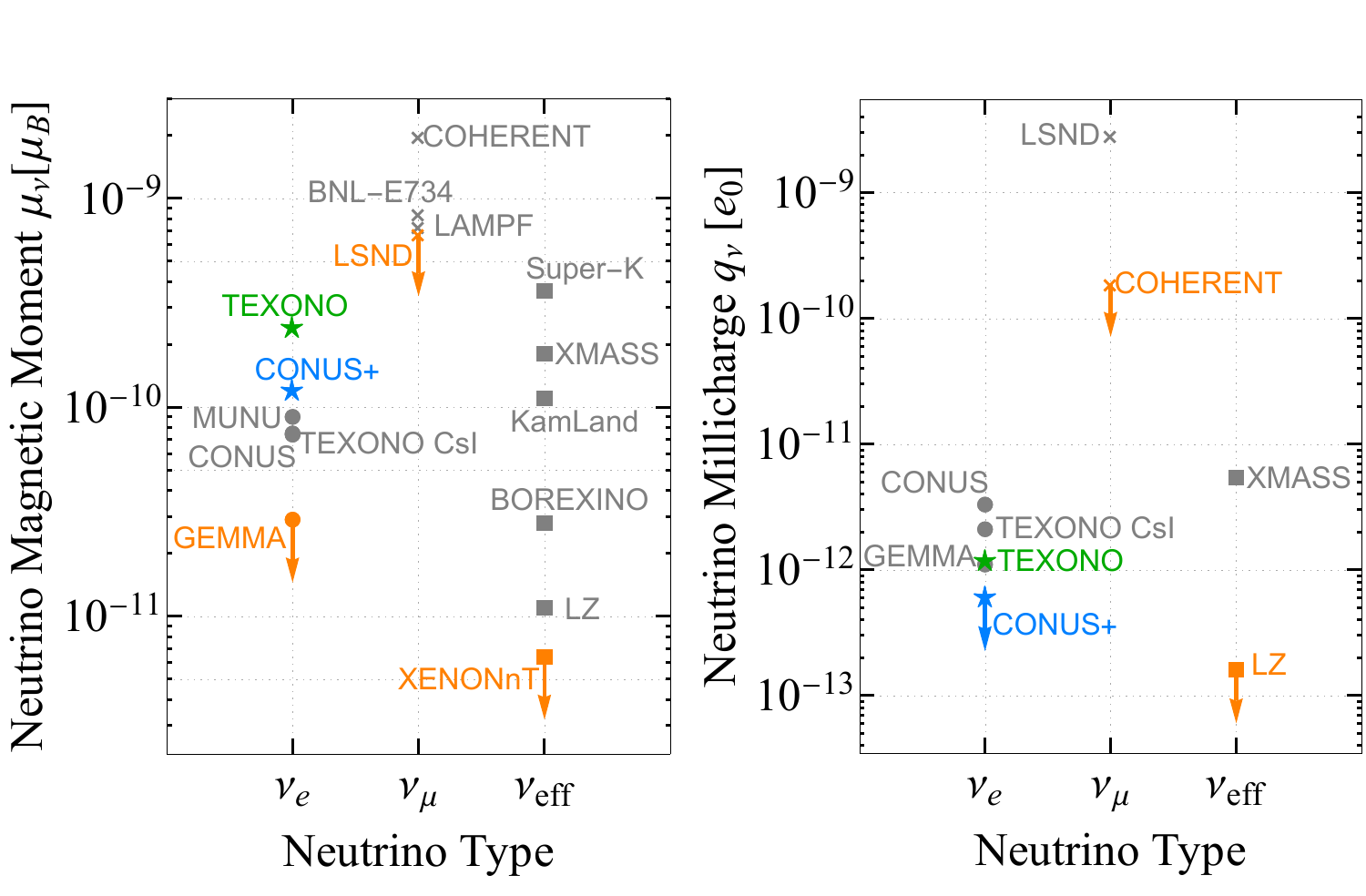}
    \caption{Summary of existing limits at 90$\%$ C.L. on the neutrino magnetic moment (left) and the neutrino millicharge (right) coming from a variety of experiments~\cite{AtzoriCorona:2022jeb,Beda:2012zz,TEXONO:2006xds,Borexino:2017fbd,Super-Kamiokande:2004wqk,cadeddudresden,MUNU:2005xnz,Allen:1992qe,Ahrens:1990fp,LSND:2001akn,Giunti:2014ixa,XMASS:2020zke,ParticleDataGroup:2024cfk,CONUS:2022qbb, XENONCollaboration:2022kmb}. The limits are divided in flavor components $\mu_{\nu_e}$ ($q_{\nu_e}$) (dots) and $\mu_{\nu_\mu}$ ($q_{\nu_\mu}$) (crosses), and also the ones on the effective magnetic moment $\mu_{\nu}^{\rm eff}$ ($q_{\nu}^{\rm eff}$) (squares) are shown. In orange, we highlighted the best limits. The results derived in this work for TEXONO and CONUS+ are shown by the green and blue stars, respectively.}
    \label{fig:MMEC}
\end{figure}

From a fit to reactor CONUS+ and TEXONO CE$\nu$NS-only data at 90\% C.L. we obtain
\begin{align}
\mu_{\nu_e}(\mathrm{CONUS+})<5.6\times 10^{-10}\mu_B, \\
\mu_{\nu_e}(\mathrm{TEXONO})<11\times 10^{-10}\mu_B,
\end{align}
while when including also the $\nu$ES channel the limits become
\begin{align}
\mu_{\nu_e}(\mathrm{CONUS+})<1.2\times 10^{-10}\mu_B, \\
\mu_{\nu_e}(\mathrm{TEXONO})<2.4\times 10^{-10}\mu_B.
\end{align}
These combined results are also shown in the left panel of Fig.~\ref{fig:MMEC} along with a compilation of other available measurements~\cite{AtzoriCorona:2022jeb}. In this context, similarly to the case of the neutrino millicharge discussed in the next section, the experimental energy threshold is the parameter that most strongly affects the sensitivity of the limits that can be set. This is why CONUS+ and TEXONO are able to place more stringent constraints than COHERENT Ar and CsI, despite the latter having significantly larger statistics. In this case, the inclusion of $\nu$GEN data, which features an energy threshold of $290\,\mathrm{eV}_{ee}$ (compared to the $200\,\mathrm{eV}_{ee}$ of TEXONO), is expected to allow for the setting of competitive and potentially interesting limits.

\subsection{Neutrino millicharge}
As already shown in many experimental and theoretical studies (for a recent review see Ref.~\cite{Giunti:2024gec}), the \cenns process is sensitive not only to the neutrino CR but also to the existence of neutrino electric charges, also known as millicharges. Indeed, even if neutrinos are considered as neutral particles, in some BSM theories, they can acquire small electric charges.
The differential \cenns cross section taking into account the contribution of the neutrino electric charges in addition to SM neutral-current weak interactions is similar to that derived for the neutrino charge radii, and is obtained by replacing the neutrino proton coupling inside the nuclear weak charge in Eq.~(\ref{eq:weakcharge}) by~\cite{Kouzakov:2017hbc,Giunti:2014ixa}
\begin{equation}
g_{V}^{p}\rightarrow g_{V}^{p}-Q_{\ell}
=
g_{V}^{p}-\dfrac{ 2 \sqrt{2} \pi \alpha }{ G_{\text{F}} q^2 }
\, q_{\nu_{\ell}}
,
\label{Qech}
\end{equation}
where $q_{\nu_{\ell}}$ is the neutrino EC. Given the extremely low momentum transfer and low-energy thresholds of reactor experiments, the $q^2$ dependence in the denominator of Eq.~(\ref{Qech}) helps to set more stringent constraints using the data of CONUS+ and TEXONO with respect to COHERENT.
The enhancement given by the presence of a neutrino electric charge becomes particularly relevant in the case of the $\nu$ES process due to the much lighter electron mass. In fact, in neutrino-electron elastic scattering, $ |q^2| = 2 m_e T_{e} $,
which is much smaller than the CE$\nu$NS $|q^2|$. Similarly to CE$\nu$NS, the contribution of EC is accounted for by replacing the neutrino vector coupling inside the cross section in Eq.~(\ref{eq:ES-cross-section}) by
$g_{V}^{\nu_{\ell}}
\to
g_{V}^{\nu_{\ell}} + Q_{\ell}$~\cite{AtzoriCorona:2022jeb}.

It is important to highlight that, while the neutrino MM cross section within the corrected FEA framework is well aligned with the predictions of {\textit{ab initio}} theories even in the sub-keV electron-recoil range, the MCRRPA cross section for a neutrino EC in this same regime exceeds the corrected FEA result by more than an order of magnitude~\cite{Chen:2014ypv,Hsieh:2019hug}. Consequently, the neutrino EC limit obtained using the FEA formalism can be considered a conservative estimate.  
Since it is well established that the EPA scheme accurately reproduces the MCRRPA cross section for a millicharged neutrino~\cite{Chen:2014ypv,Hsieh:2019hug}, we adopt the EPA formalism as an improvement over the FEA approach which provides a more precise description of the interaction. This refined framework is expected to impose more stringent constraints on the neutrino millicharge. In particular, the EPA cross section for a millicharged ultrarelativistic particle is given by~\cite{Chen:2014ypv,Hsieh:2019hug, AtzoriCorona:2022jeb}
\begin{equation}
    \dfrac{d\sigma_{\nu_\ell}}{d T_\text{e}}\Big\vert_{\rm{EPA}}^{\rm{EC}}=\frac{2\alpha}{\pi}\frac{\sigma_\gamma(T_e)}{T_e}\ln\left[\frac{E_\nu}{m_\nu}\right]q_{\nu_\ell}^2
    \label{eq:EPA},
\end{equation}
where $m_\nu$ represents the neutrino mass, set conservatively to $1\;\rm{eV}$~\cite{ParticleDataGroup:2024cfk}, while $\sigma_\gamma(T_e)$ denotes the photoelectric cross section for a real photon, which can be experimentally measured~\cite{HENKE1993181} for Ge. Moreover, in this case we enlarge the systematic contribution to $\sigma_\eta=0.2$ in order to take into account further uncertainties on the EPA approach~\cite{Chen:2014dsa}.
From Eq.~(\ref{eq:EPA}), it is evident that, unlike the FEA approximation, the cross section in the EPA framework remains unaffected by the sign of the electric charge.

\begin{figure}[t]
    \centering
    \includegraphics[width=0.95\linewidth]{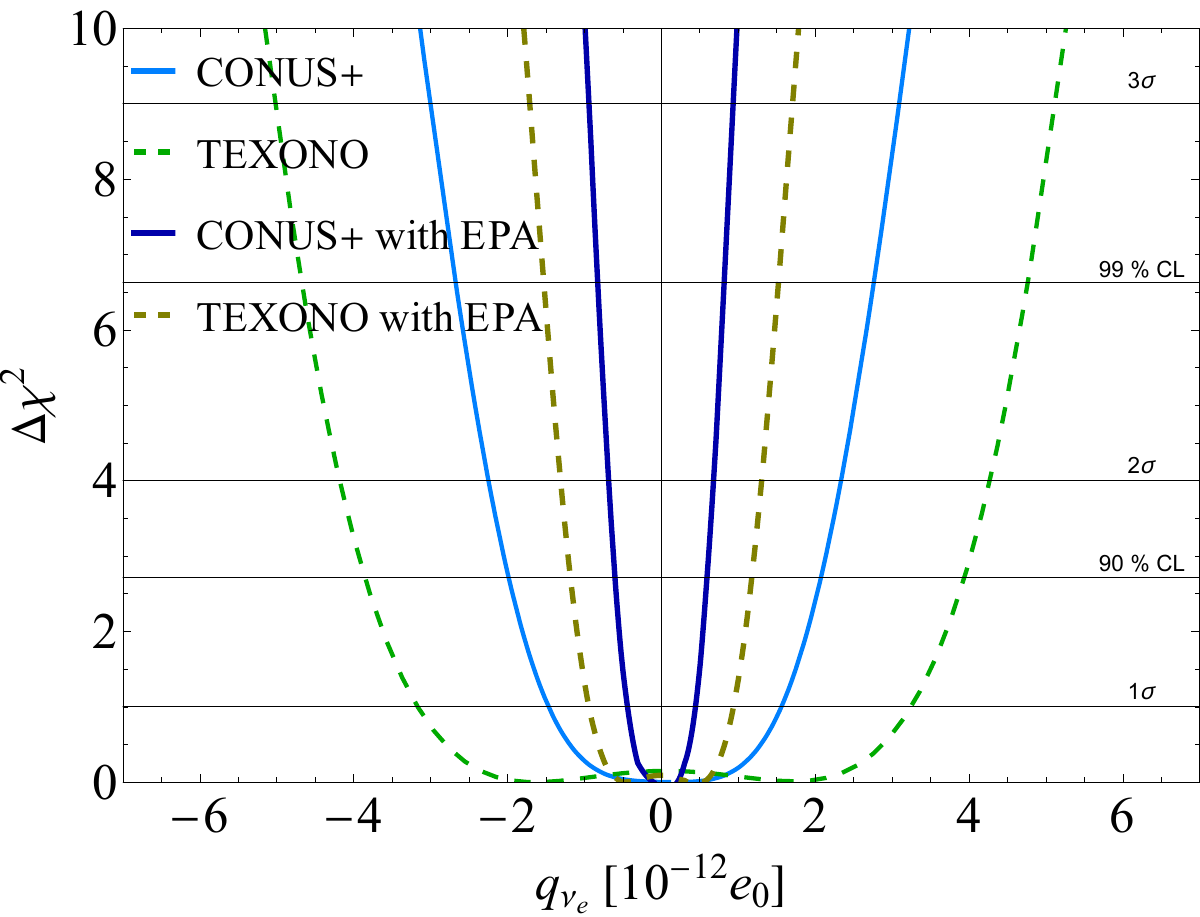}
    \caption{Marginal $\Delta\chi^2$'s for the electron neutrino EC, $q_{\nu_e}$, using TEXONO (dashed green lines) and CONUS+ (solid blue lines) datasets considering both the CE$\nu$NS and $\nu$ES channels. The most stringent limits are obtained with the EPA approach (darker lines) while the weaker limits are obtained using FEA (lighter lines).}
    \label{fig:ES}
\end{figure}

Since the inclusion of the $\nu$ES channel improves the limits by about 3 orders of magnitude, in Fig.~\ref{fig:ES} we show the constraints on the neutrino EC obtained fitting the TEXONO and CONUS+ data accounting also for the $\nu$ES channel using FEA and EPA approaches. It is clear that the use of EPA results in a further improvement of about a factor 3 on EC, which at 90\% C.L. reads
\begin{align}
    &{\rm{EPA\, CONUS+:}}\,\,\,\,-0.6 < q_{\nu_e} [10^{-12}e_0]<0.6,\\
    &
    {\rm{EPA\, TEXONO:}}\,\,\,\,-1.2 < q_{\nu_e}[10^{-12}e_0]<1.2.
\end{align}

These results are summarized in the right panel of Fig.~\ref{fig:MMEC}, from which it is possible to see that CONUS+ is able to reach the best limit on the electron neutrino millicharge,  $q_{\nu_e}$,  while TEXONO still achieves a competitive constraint.

\subsection{Neutrino NSI and light mediators}
\begin{figure}[h]
    \centering
    \subfigure[]{\label{fig:ptNSI}
    \includegraphics[width=0.9\linewidth]{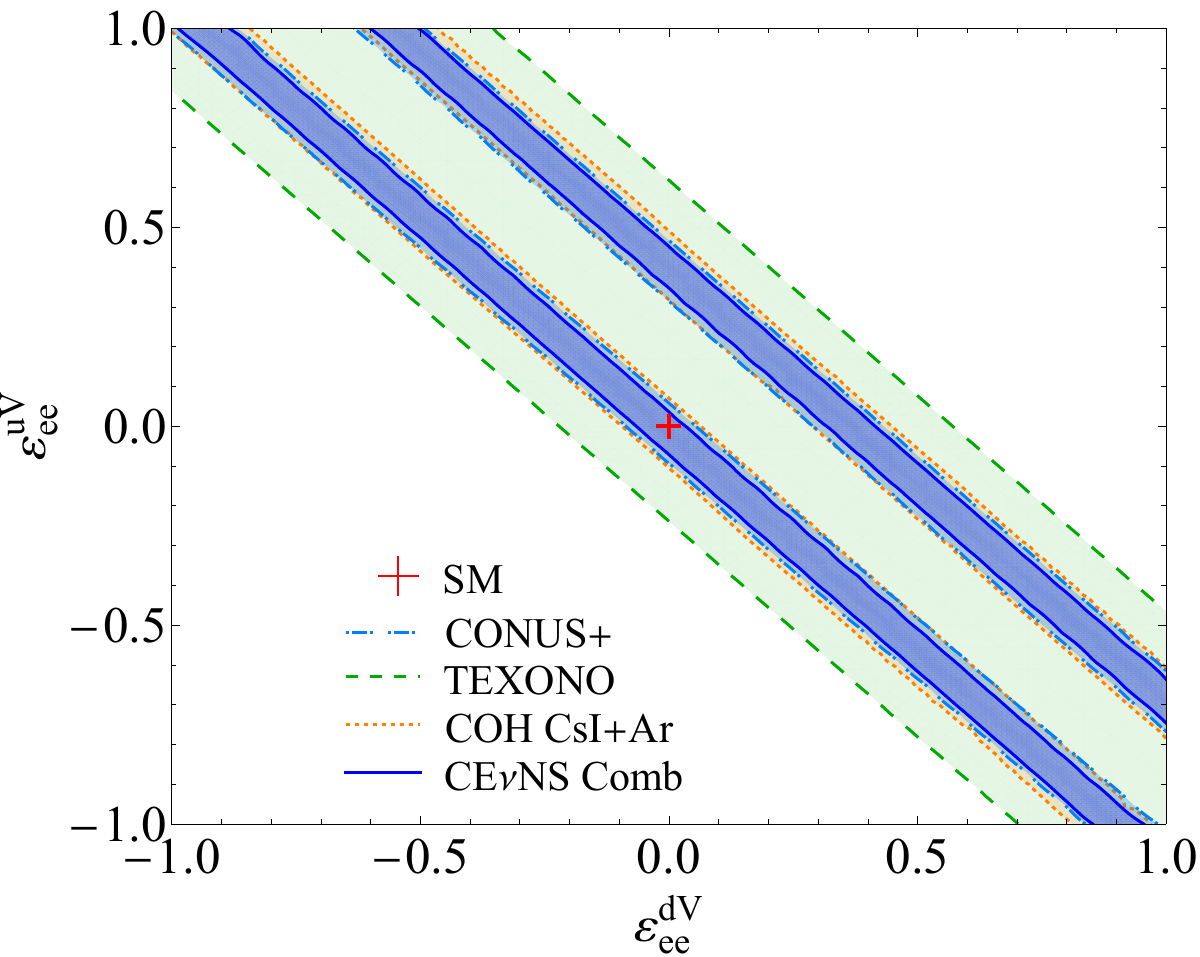}}
    \subfigure[]{\label{fig:ptLM}
    \includegraphics[width=0.95\linewidth]{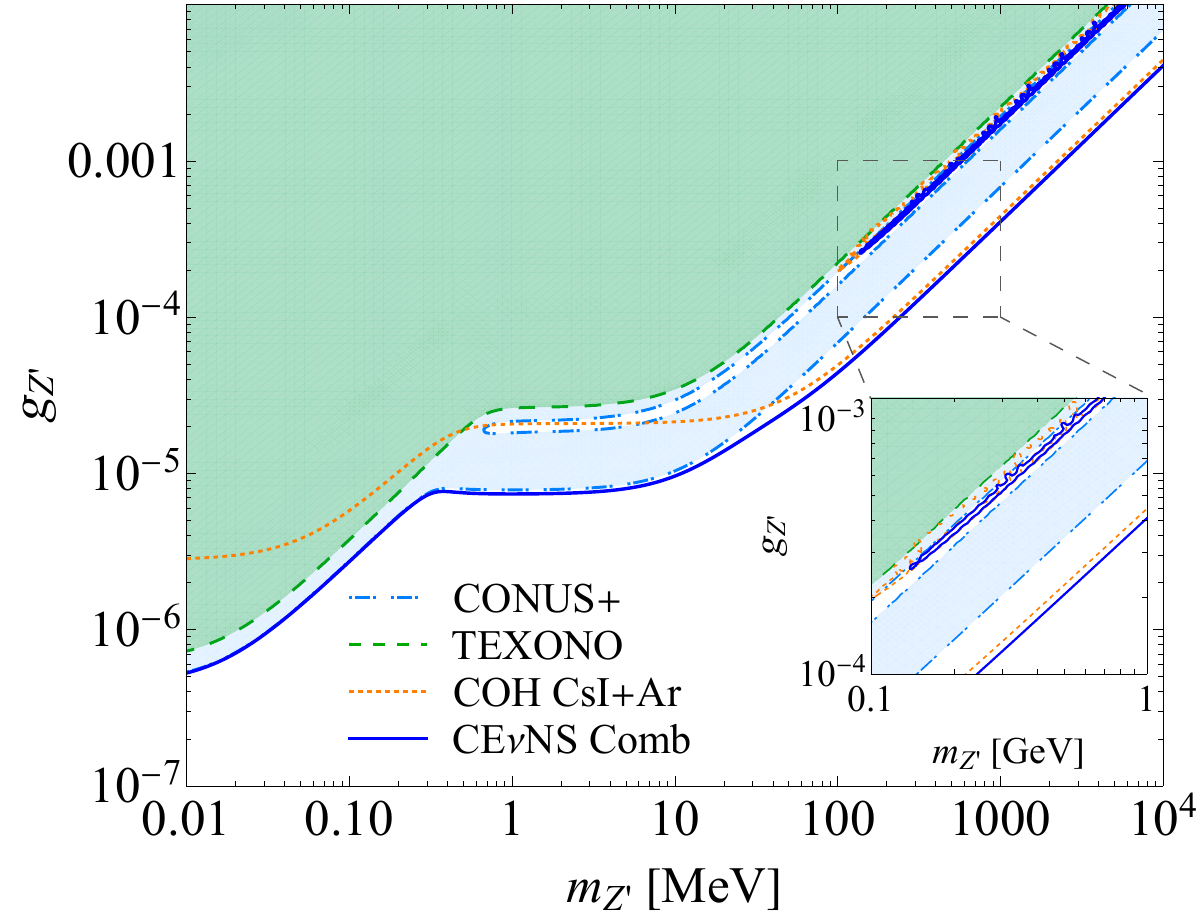}}
    \caption{Constraints on flavor-preserving NSI (a) and the universal light mediator $Z'$ (b) at 90\% CL. The TEXONO (green dashed contour) and the CONUS+ (dot-dashed light blue contour) are shown along with the combined COHERENT CsI+Ar result (dotted orange contour) and their combination (solid blue contour). The red cross indicates the SM. In the inset (b), we present an enlargement of the contour around the degeneracy strip.}
    \label{fig:enter-label}
\end{figure}

In the presence of a new massive vector mediator that couples to SM leptons and quarks, the SM cross section presented in Eq.~(\ref{eq:cexsec}) is modified. Assuming that the neutrino does not change flavor, the effect of such a nonstandard interaction (NSI) is generically described by the effective four-fermion
interaction Lagrangian~\cite{Giunti:2019xpr, Coloma:2023ixt}
\begin{equation}
\mathcal{L}_{\text{NSI}}^{\text{NC}}
=
- 2 \sqrt{2} G_{\text{F}}\sum_{\ell=e,\mu}
\left( \overline{\nu_{\ell L}} \gamma^{\rho} \nu_{\ell L} \right)
\sum_{f=u,d}
\varepsilon_{\ell\ell}^{fV}
\left( \overline{f} \gamma_{\rho} f \right)
\, ,
\label{lagrangian}
\end{equation}
where $\nu_{\ell \text{L}}$ and $f$ represent the neutrino and the fermion fields, respectively.
The parameters
$\varepsilon_{\ell\ell}^{fV}$, where $f=u,d$ stands for the flavor of the quark and $\ell=e,\mu$ is the neutrino flavor\footnote{We consider only the first generation of quarks since they are the only ones contained in nuclei and electronic and muonic neutrinos in the case of COHERENT measurements and only electronic neutrinos for reactor experiments.}, describe the size of nonstandard interactions relative to standard neutral-current weak interactions. Thus, the full \cenns cross sections is obtained by replacing the nuclear weak charge in Eq.~(\ref{eq:weakcharge}) by~\cite{Cadeddu:2020nbr,AtzoriCorona:2022moj}
\begin{eqnarray}
Q_{\ell,\mathrm{NSI}}^{V}
&=&
\left( g_{V}^{p}(\nu_{\ell}) + 2 \varepsilon_{\ell\ell}^{uV} + \varepsilon_{\ell\ell}^{dV} \right)
Z
F_{Z}(|\vet{q}|^2)
+ \nonumber \\ 
&+&\left( g_{V}^{n} + \varepsilon_{\ell\ell}^{uV} + 2 \varepsilon_{\ell\ell}^{dV} \right)
N
F_{N}(|\vet{q}|^2).
\label{Qalpha2}
\end{eqnarray}
It is clearly visible that the contributions from a new NSI vector mediator may produce an interference with the SM.
In particular, in this work, we consider the simplified scenario with only two nonzero NSI parameters, $\varepsilon_{ee}^{uV}$ and $\varepsilon_{ee}^{dV}$, also known as flavor-preserving scenario, which involves only the electronic neutrino flavor.\\
The constraints obtained using TEXONO and CONUS+ on the latter parameters are shown in Fig.~\ref{fig:ptNSI} at 90\% C.L. and compared to the ones from COHERENT CsI and Ar data. The contours present two diagonal allowed strips which correspond to possible degeneracies inside the nuclear weak charge. The blue contour represents the result of the combined analysis of the aforementioned \cenns data. Indeed, by combining \cenns data on different nuclei, it is possible to significantly reduce the allowed parameter space, profiting from the different neutron and proton number combinations inside the nuclear weak charge in Eq.~(\ref{Qalpha2}). Considering that both COHERENT Ge and $\nu$GEN employ the same target nuclei as TEXONO and CONUS+, any gain in sensitivity is expected to solely come from the additional statitics.\\

In the scenario in which NSIs are mediated by a light vector boson, usually referred to as $Z'$, the NSI parameters take the form of a propagator~\cite{Cadeddu:2020nbr,AtzoriCorona:2022moj},
\begin{equation}
    \epsilon_{\ell \ell}^{fV}=\dfrac{g_{Z'}^2\,Q'_\ell Q'_f}{\sqrt{2}G_F\, (|\vec{q}|^2+m_{Z'}^2)},
\end{equation}
where $m_{Z'}$ and $g_{Z'}$ represent the gauge boson mass and coupling, respectively, and $Q'$ are the charges under the new gauge symmetry $U(1)'$.

Therefore, the \cenns cross section in the presence of a light vector mediator can be retrieved by rewriting the NSI parameter in the form of the light $Z'$ propagator inside the nuclear weak charge in Eq.~(\ref{Qalpha2}).
To highlight the capabilities of reactor \cenns experiments, we have considered a simple model in which the novel $Z'$ boson couples universally to all SM fermions~\cite{Liao:2017uzy,Papoulias:2017qdn,Papoulias:2019txv,Cadeddu:2020nbr,Bertuzzo:2021opb,AtzoriCorona:2022moj}, {\it{i.e.}} i.e. the universal model.
We set the charges to be $Q'_\ell \equiv Q'_f = 1$, and the coupling becomes the same for all the fermions.
Given that the $Z'$ propagator depends on the experimental momentum transfer, including also the neutrino-electron scattering process allows one to extend the constraints toward lighter $Z'$ masses. In the case of $\nu$ES, the cross section for the light mediator contribution can be obtained by substituting the neutrino-electron vector coupling by~\cite{DeRomeri:2024dbv,DeRomeri:2025csu}
\begin{equation}
    g_V^{\nu_\ell}\rightarrow g_V^{\nu_\ell}+\dfrac{g_{Z'}^2\,Q'_\ell Q'_e}{\sqrt{2}G_F\, (|\vec{q}|^2+m_{Z'}^2)}\, .
\end{equation}
The constraints obtained on the universal light mediator model are presented in Fig.~\ref{fig:ptLM}, where the comparison with those from COHERENT CsI and Ar data is also shown. It is evident that the contribution given by the $\nu$ES process enhances the sensitivity in the low mass region. Moreover, both COHERENT and CONUS+ present an unconstrained diagonal strip which occurs when the $Z'$ parameters produce a degenerate cross section with respect to the SM one, as explained in Refs.~\cite{Cadeddu:2020nbr,AtzoriCorona:2022moj}. The location of such a degeneracy depends on the ratio between the neutron number and the atomic number, and thus by combining \cenns data on different nuclei, it is possible to significantly reduce such degeneracy. This can be better appreciated in the inset in Fig.~\ref{fig:ptLM} where a zoom of the degeneracy strip parameter region is shown. The inclusion of additional data, such as the COHERENT Ge and $\nu$GEN data, could allow further reduction of the degeneracy strip.

\section{Conclusions}\label{sec:conclusions}

The recent CE$\nu$NS observations by the TEXONO and CONUS+ Collaborations mark a significant step forward in the study of neutrino interactions at low energies. Their results confirm the validity of the Lindhard quenching model for germanium detectors and confirm the \cenns observation in the low-energy regime. These findings reinforce the role of reactor-based experiments as a complementary approach to high-energy \cenns studies at the SNS.\\

By leveraging the data from CONUS+ and the constraints from TEXONO, we have explored their implications for SM and BSM physics. In particular, we have provided novel measurements of the weak mixing angle, in good agreement with the SM prediction, and improved sensitivities to neutrino electromagnetic properties, such as charge radius, electric millicharge, and magnetic moment. In the last two cases, we have examined the role of elastic neutrino-electron scattering, which becomes particularly relevant when considering these BSM scenarios. Thanks to the low-energy threshold and the dependence on only one neutrino flavor,  TEXONO and CONUS+ data lead to very competitive limits on these quantities, and in particular when using the EPA approach to model the neutrino-electron interaction, we obtain the most stringent limit on the electron neutrino millicharge.
Additionally, we have set updated constraints on nonstandard interactions and the existence of light mediators. The combination of different target materials with varying proton and neutron numbers, along with the minimal dependence of reactor data on nuclear form factors, enables a significant improvement over existing limits. Notably, in the light mediator case, this permits the reduction of the degeneracy strip in the parameter space.\\

The combined results from reactor-based and SNS \cenns measurements establish a robust framework for testing the Standard Model and beyond with intriguing precision. Future improvements in statistics and further experimental efforts will be crucial for refining these measurements and probing new physics with even greater accuracy.\\

\begin{acknowledgements}
The authors gratefully acknowledge C. Buck, M. Lindner, and W. Maneschg for their valuable support and the up-to-date information provided on behalf of the CONUS Collaboration, which was essential for the accurate interpretation of the CONUS+ data. 
The authors are also thankful to V. De Romeri, D. K. Papoulias and G. Sanchez Garcia for the fruitful discussions on the CONUS+ data analysis.
\end{acknowledgements}

\bibliography{ref}

\end{document}